\documentclass[12pt]{article}
\usepackage{scicite}
\usepackage{times}
\usepackage{amsmath}
\usepackage{graphicx}
\usepackage{float}
\newcommand{\bbra}{\left\langle}
\newcommand{\kket}{\right\rangle}

\newcommand{\ket}[1]{\left. |#1 \right  \rangle}
\newenvironment{sciabstract}{%
\begin{quote} \bf}
{\end{quote}}

\title{High Temperature Superconductivity: A Simple Model Exploiting Hydrogen Bonds}

\author
{Daniel Kaplan,$^{1\ast}$ Yoseph Imry,$^{1}$\\
\\
\normalsize{$^{1}$Dept. of Condensed Matter Physics, Faculty of Physics, Weizmann Institute of Science}\\
\normalsize{Rehovot, 76100, Israel}\\
\\
\normalsize{$^\ast$To whom correspondence should be addressed; E-mail:  daniel.kaplan@weizmann.ac.il}
}

\date{}

\begin{document} 


\baselineskip24pt


\maketitle


\begin{sciabstract}
Lately, there has been much interest in high temperature superconductors, and more recently hydrogen-based superconductors. This work offers a simple model which explains the behavior of the superconducting gap based on BCS theory, and reproduces most effects seen in experiments, including the isotope effect and $T_c$ enhancement as a function of pressure. We show that this is due to a combination of the factors appearing in the gap equation: the matrix element between the proton states, and the level splitting of the proton.
\end{sciabstract}

% In setting up this template for *Science* papers, we've used both
% the \section* command and the \paragraph* command for topical
% divisions.  Which you use will of course depend on the type of paper
% you're writing.  Review Articles tend to have displayed headings, for
% which \section* is more appropriate; Research Articles, when they have
% formal topical divisions at all, tend to signal them with bold text
% that runs into the paragraph, for which \paragraph* is the right
% choice.  Either way, use the asterisk (*) modifier, as shown, to

\section*{Introduction}

Recently, there has been a surge of interest concerning the discovery of high temperature superconductivity in hydrogen compounds \cite{Errea2016,Drozdov2014,Flores-Livas2015}. These showed that metallization and superconductivity in such materials can be obtained by changing the pressure of the system. Furthermore, the discovery of the inverse isotope effect in hydrides (notably, PdH) \cite{Skoskiewicz1974} helped fuel interest in these materials. Historically, arguments have been put forward that the maximal value of $T_c$ in BCS theory is roughly $T_c \sim 30K$\cite{Bardeen1957}; but since then higher values of $T_c$ were demonstrated, e.g. in the discovery of the cuperates, pnictides and $MgB_2$\cite{Anderson2013}. Suggested first by N. Ashcroft \cite{Ashcroft1968}, metallic hydrogen (and hydrogen compounds) can be a platform for high temperature superconductivity. This is accomplished by maximizing the $N_0 U$ parameter in the BCS gap equation, while increasing the energy transfer to the phonon $\Delta E$. In this paper, we present a model which accounts for these properties in hydrogen sulfide, and gives a qualitative explanation for the enhancement of $T_c$, the isotope effect in these compounds, and the reason for pressure dependence in enhancing (or weakening) $T_c$.

\section*{The Harmonic oscillator case}
We first consider the emergence of a superconducting gap by assuming that electrons interact with a harmonic oscillator, 
\begin{align}
H = \frac{p^2}{2M} + \frac{M \omega_0^2 x^2}{2}
\end{align}
With well-known solutions for wavefunctions and eigen-energies. 
We consider the usual virtual transitions between the ground-state and a phonon-excited state, which have an energy difference $\Delta E = \hbar \omega_0$, and another eletron de-excites that state. Following \cite{DeGennes1996} for the calculation of the full attractive interaction matrix element.
\begin{align}
\bbra 1 |H_{int} |2 \kket = 2 \frac{|W|^2_q}{\hbar} \left(\frac{\omega_q}{\omega^2-\omega_q^2}\right)
\label{f_mat_elem}
\end{align}
Here the states $\ket 1$, $\ket{2}$ correspond to two Cooper pairs. One with incoming momentum pairs $(k, -k)$, and the other $(k', -k')$. $\vec{q} = \vec{k}-\vec{k}'$. $W_q$ is the relevant electron-phonon matrix element, for the emission or absorption of a phonon with wave-vector $\vec{q}$. $\omega_q$ is the energy of the phonon, and $\omega = |E(1) - E(2)|$ is the energy difference between states (1 and 2, respectively). We will consider the particle occupying the H.O. as ``the proton". In the lattice, each energy level becomes a narrow band, with states characterized by a wave-vector $\vec{q}$.
\subsection*{Electron-proton coupling}
We explore two forms of this term. For simplicity, we show this in 1D but the approach is easily generalized to 3D. Firstly, we approximate the electron-proton interaction by a delta function, with the form $V_{int} = -g \delta(x-y)$,
such that $x$ is the coordinate of the oscillator, and $y$ is the coordinate of the electron. Suppose that an electron excites this H.O. from its ground state to the first excited state. The initial state of the electron,$\ket{i} = \frac{1}{\sqrt{2 \pi}} \exp(-iky)$
and the final state (after interacting with the oscillator),$\ket{f} = \frac{1}{\sqrt{2 \pi}} \exp(-ik'y)$. All in all the total state of the system is expressed by the notation $\ket{{0, i}}$ (where the $0$ refers to the ground state of the oscillator) and $\ket{{1, f}}$, where the $1$ now indicates the first excited state. Then the matrix element is,
\begin{align}
\begin{matrix}
\displaystyle W_q = \bbra 0, i | V_{int}  |1, f \kket = -\frac{g}{2\pi} \iint dx dy \psi_0^{*}(x) e^{\left(-i(k-k')y\right)} \delta(x-y)\psi_1(x) = \\ \displaystyle -\frac{g}{2\pi}\int dx \psi_0^{*}(0)\psi_1(x) e^{-iqx}
\label{mat_elem}
\end{matrix}
\end{align}
Where we denoted $k-k' = q$, i.e., the momentum transfer to the phonon. $\psi_0, \psi_1$ are the ground-state, first excited state wavefunctions, respectively. For a harmonic oscillator this result is known exactly. We have,
\begin{align}
W_q = \frac{i g q a}{2 \pi } e^{-\frac{q^2 a^2}{4}}
\label{ho_elem}
\end{align}
Where $a^2 = \frac{\hbar}{M \omega_0}$ is the length scale of the oscillator. In our case, $qa \ll 1$, and a reasonable value for $g$ above is $g \sim 2 eV \AA$.
We also consider a screened-Coulomb interaction. For this form, $V_{int} = -\frac{Z e^2}{4 \pi} \frac{e^{-k_s r}}{r}$,
where $k_s$ is the screening length, the matrix element of Eq. \ref{mat_elem} becomes,
\begin{align}
\bbra 0, i | V_{int}  |1, f \kket =  -\frac{Z e^2}{8\pi^2}\int dx \psi_0^{*}(0)\psi_1(x) \frac{e^{-iqx}}{q^2 + k_s^2}
\end{align}
For H\textsubscript{2}S, in the metallic phase, $k_s \sim 1 \frac{1}{\AA}$ \cite{Havriliak1955}.
\subsection*{The BCS superconducting gap}
We focus on a particular longitudinal phonon mode, which, by work in \cite{Drozdov2014}, has been shown to be particularly dominant in these materials. Consider the matrix element of Eq. \ref{f_mat_elem}. Substituting $W_q$ from Eq. \ref{ho_elem} one finds,
\begin{align}
U = \bbra 1 | H_{int} | 2 \kket = \frac{g^2 q^2 e^{-\frac{q^2 \hbar }{2 m \omega_0 }}}{2 m \omega _0}\times\frac{\omega_q}{\omega^2-\omega_q^2}
\end{align}
We make the simplifying assumption that the processes we're interested in occur at $\omega = 0$. We base this on the fact the probability of inelastic processes where $\omega \neq 0$ is relatively small (which is implied by the $qa \ll 1$ limit)\cite{DeGennes1996}. Finally, we are left with,
\begin{align}
U = -2 \frac{g^2 q^2 a^2 e^{-\frac{q^2 a^2}{2}}}{\hbar \omega_0}
\end{align}
This form of the coupling is also found in \cite{Imry1969}. We then replaced $\omega_q = \omega_0$ since the phonon in this case is represented by the H.O, and the relevant branch has a very small dispersion in $\vec{q}$. The plot below shows $U$ as a function of $\omega_0$ for a given $q$. Typical $q$ is $q \sim 0.1 \frac{1}{A}$. $g \sim 2 eV$, M is the proton mass.. The relevant dimensionless parameter is the BCS interaction parameter, which we take to be $N_0 U$, where $N_0$ is the electronic density at the Fermi energy. For $H_2S$, we have $N_0 \sim 0.3 eV^{-1}$ \cite{Kudryashov2017}.
\begin{figure}[H]
\centering
\includegraphics[width=0.8\textwidth]{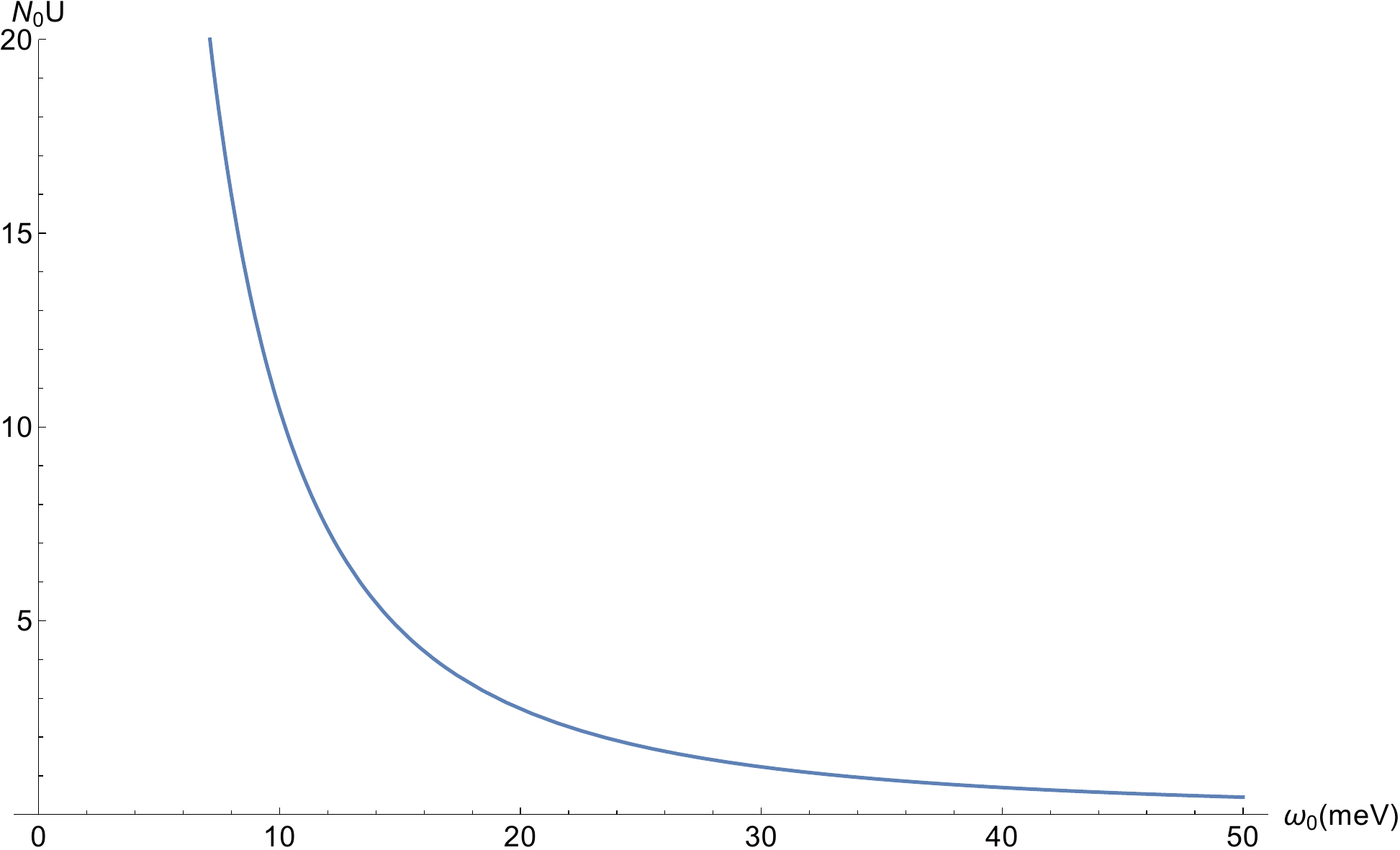}
\caption{$N_0 |U|$ as a function of $\omega_0$}
\label{mat_elem_harm}
\end{figure}
The BCS gap has the form \cite{Bardeen1955}\cite{DeGennes1996},
\begin{align}
\Delta = \hbar \omega_0 e^{-\frac{1}{N_0 |U|}}
\label{gap_eq}
\end{align}
Note that the prefactor to the exponential is an increasing function of $\omega_0$, whereas the matrix element, per Fig. \ref{mat_elem_harm} is a decreasing one. Clearly, a maximum might exist, This is shown below.
\begin{figure}
\centering
\includegraphics[width=0.8\textwidth]{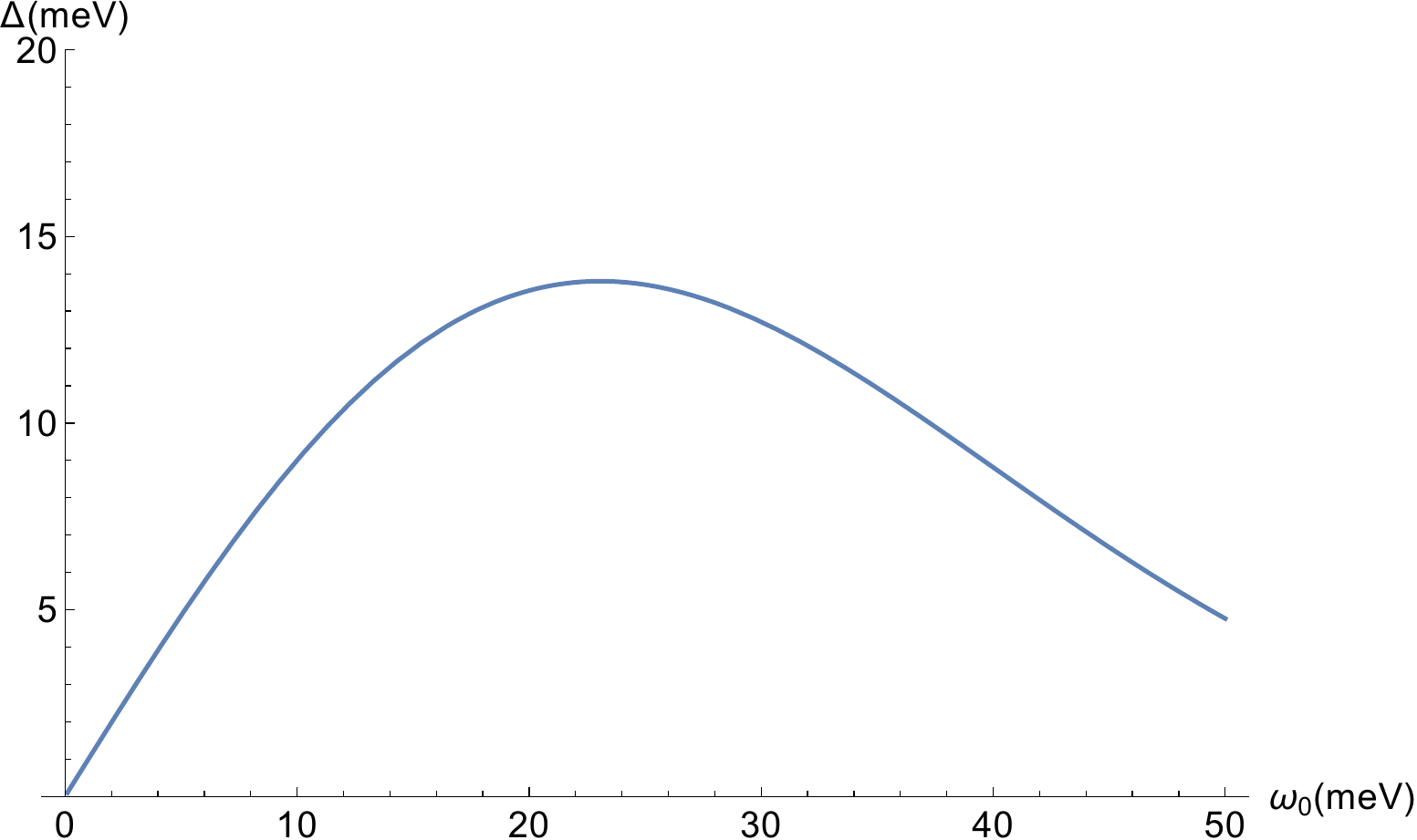}
\caption{$\Delta $ as a function of $\omega_0$, for $q \sim 0.1 \frac{1}{A}$, according to Eq. \ref{gap_eq}}
\label{delta_omega}
\end{figure}
Using results of the gap equation at finite temperatures (see below), the maximum in Fig. \ref{delta_omega} appears at $\omega_0 = 22 meV$, and corresponds to $T_c \approx 92K$.
\subsection*{The Isotope effect}
Having found $\Delta$ it is possible to investigate the dependence of the gap on the Mass of the proton. We calculate,
\begin{align}
\frac{1}{\Delta} \frac{d \Delta}{d M} \sim -\frac{\omega_0^2}{q^2}\left(1- \frac{q^2 a^2}{2}\right)
\end{align}
Recall that $a^2 \sim \frac{1}{M \omega_0}$, this shows, as suggested by Fig. \ref{delta_omega}, one may obtain either a positive, or negative isotope effect, depending on the sign of $1 - q^2 a^2 /2$. In the range $qa \ll 1$ the gap changes negatively (i.e. decreases) with mass. This is confirmed by \cite{Drozdov2014, Errea2016}, and experimentally observed.
\subsection*{Gap equation at finite temperature}
For completeness, we include the gap equation at finite temperature which permits an accurate determination of $T_c$. BCS theory puts $T_c \sim \Delta$ (up to a numerical factor of order 1) \cite{Bardeen1955}. The BCS result is (using our simplified approach),
\begin{align}
1 = N_0 U \int \limits_0^{\Delta E} \frac{d E}{E^2 + \Delta^2} \left(1- 2f(\sqrt{E^2+\Delta^2})\right)
\end{align}
Where $f$ is the Fermi-Dirac distribution. At the transition temperature $T=T_c$, the gap $\Delta = 0$, yielding,
\begin{align}
T_c \sim \frac{\Delta}{1.75}
\end{align}
\section*{A simple new model of the superconducting transition in $H_2S$}
Numerical as well as experimental work on the structure of $H_2 S$ have confirmed that the material experiences a transition of crystalline-structure upon becoming superconducting \cite{Ge2016}. 
In the $Im\bar{3}m$ phase of hydrogen sulfide, shown below, the hydrogen bond between $S$ atoms is symmetrized such that the body-centered phase allows for superconductivity. 
\begin{figure}[H]
\centering
\includegraphics[width=0.2\textwidth]{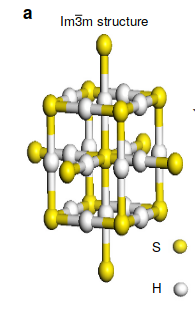}
\caption{The body-centered structure of $H_3 S$, taken from \cite{Ge2016}}
\label{h_s_bond}
\end{figure}
We model the potential seen by the proton using a sum of two Lennard Jones (L-J) potentials \\ $V = 4 \epsilon \left(\left(\frac{\sigma}{r}\right)^6-\left(\frac{\sigma}{r}\right)^{12}\right)$.
Consider placing two $S$ atoms at co-linear points separated by $x_0$ in space (for simplicity, the dynamics are on the $x$ axis). This was observed, experimentally (in hydrogen-bonded compounds) \cite{Imry1965}. Then, the potential an $H$ atom experiences is,
\begin{align} 
V = 4 \epsilon \left(\left(\frac{\sigma}{(x-x_0)}\right)^6-\left(\frac{\sigma}{(x-x_0)}\right)^{12}\right)+4 \epsilon \left(\left(\frac{\sigma}{(x+x_0)}\right)^6-\left(\frac{\sigma}{(x+x_0)}\right)^{12}\right)
\label{act_pot}
\end{align}
The parameter governing the form of the potential is then $x_0$ -- such that $2x_0$ is the separation of $S$ atoms. Below are several forms of the potential for different values of $x_0$.
\begin{figure}[H]
\centering
\includegraphics[width=0.8\textwidth]{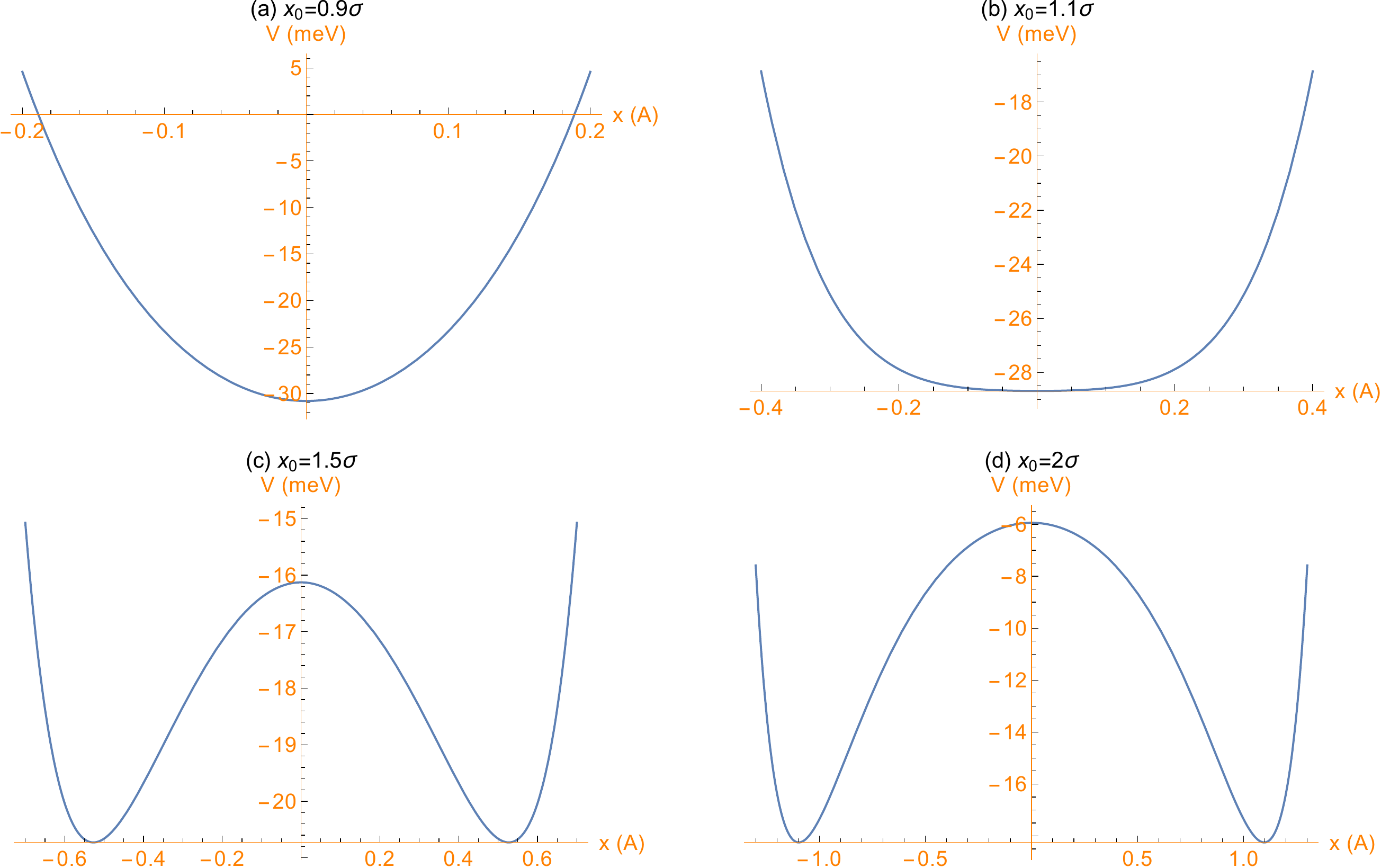}
\caption{Various forms of the potential in Eq. \ref{act_pot}. Note the transition from single-well to double-well, occurring at $x_0 \approx 1.2 \sigma$}
\label{pot_fig}
\end{figure}
Typical values for $H_2 S$ are $\epsilon \sim 300 K$,$\sigma \sim 2.0 A$ \cite{Galliero2008}. After diagonalizing the Hamiltonian to find the energy levels and eigenfunctions, required by the procedure in Eq. \ref{mat_elem}. The figure below shows the first two levels' splitting, as a function of $x_0$, 
\begin{figure}[H]
\centering
\includegraphics[width=0.8\textwidth]{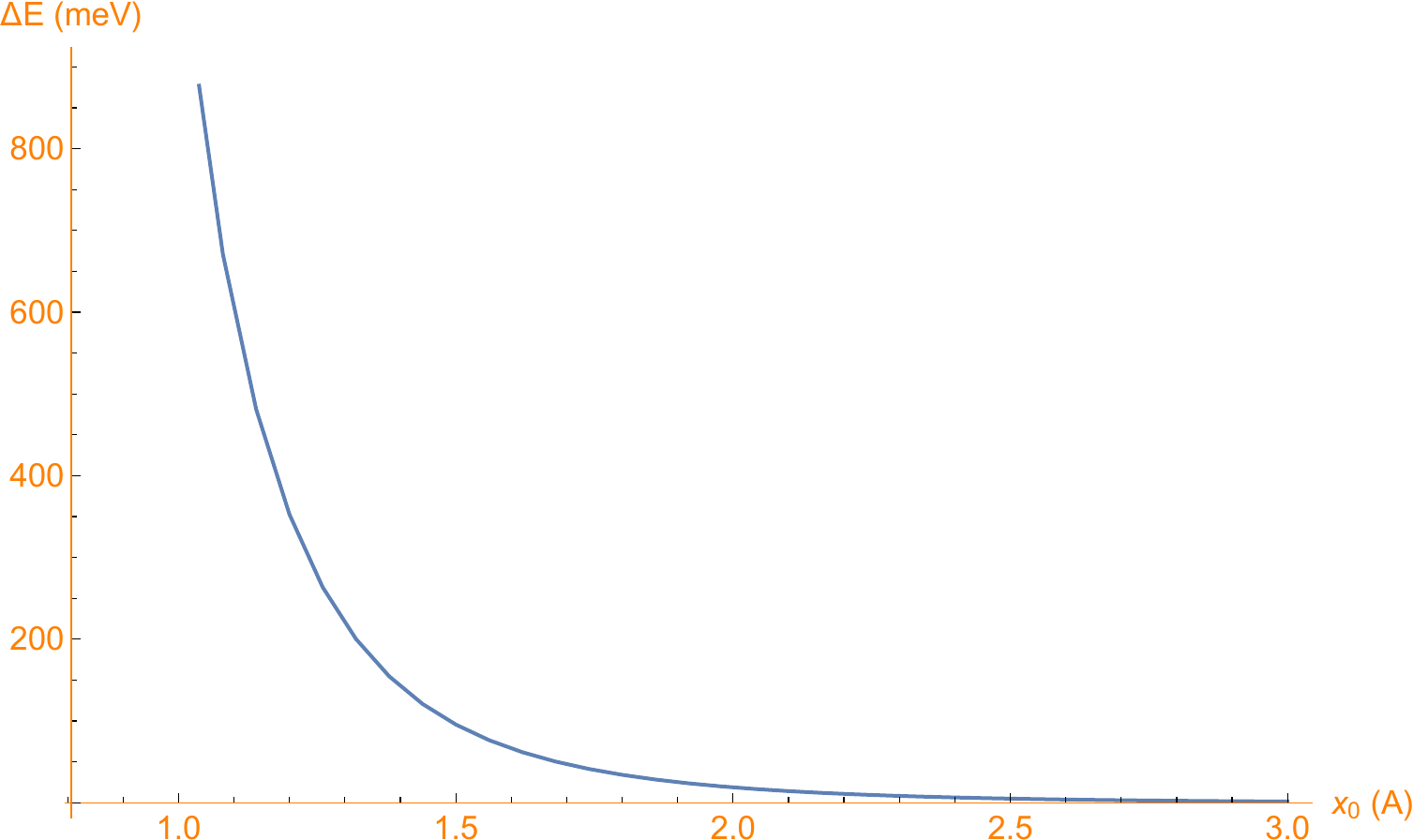}
\caption{$\Delta E_{12}$ (energy difference between ground state and first excited state). $\sigma \sim 2.0 A$, $\epsilon \sim 300K$}
\label{deltaE}
\end{figure}
Using Eqs. \ref{f_mat_elem},\ref{gap_eq}, we can now obtain the superconducting gap. We replace above, in Eq. \ref{gap_eq}, $\omega_0$ with the energy difference $\Delta E \equiv \Delta E (x_0)$. The usual solution is then,
\begin{align}
\Delta = (\Delta E) e^{-\frac{1}{N_0 U}}
\label{true_gap}
\end{align}
Where $N_0 \sim 0.3 \frac{1}{eV}$. Finally, the plot below shows all the relevant factors that enter Eq. \ref{true_gap}. The red line depicts the energy difference $\Delta E$. The brown line is the matrix element squared as it appears in Eq. \ref{mat_elem}. The combination of these, detailed above gives the gap, which is plotted in blue.
\begin{figure}[H]
\centering
\includegraphics[width=0.8\textwidth]{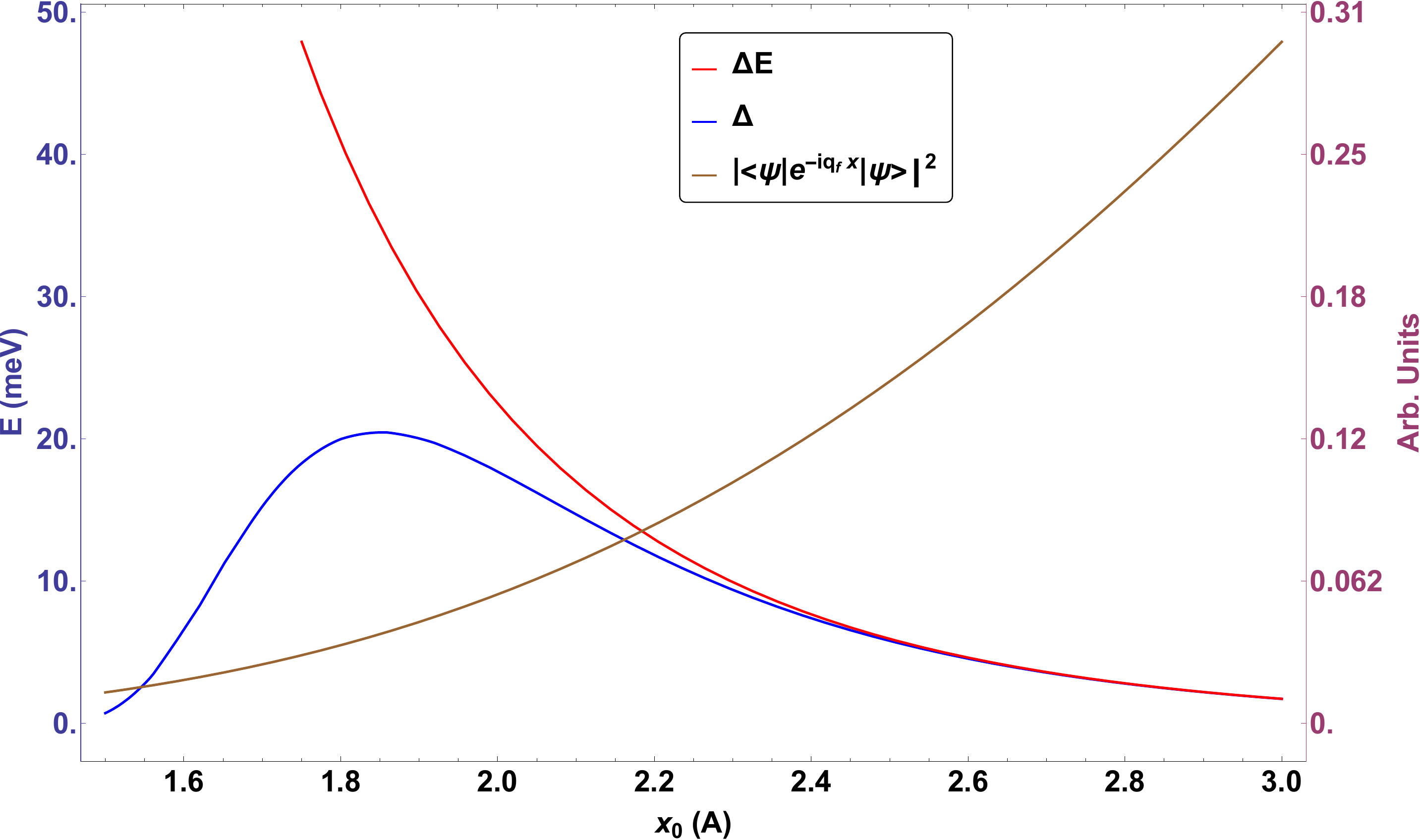}
\caption{Numerical results for the L-J potential. Red is the energy difference between the first two levels, brown is the electron-proton interaction matrix element squared, blue is the resultant gap}
\label{full_plot}
\end{figure}
The figure above clearly illustrates that $\Delta$ possesses a maximum. This is obtained at $x_0 \approx 1.82 A$, with a value of $\Delta \approx 22 meV$, giving a $T_c \approx 150K$. This is a considerable enhancement over the harmonic case (shown before), and suggests that (relatively) high $T_c$ can be obtained in this model. \textbf{This follows from the large matrix element, resulting from the separation of the two minima}.
The change in the separation may be caused by pressure on the system which modifies the positions of the $S$ atoms. 
\section*{Conclusions}
This work demonstrates qualitatively that $T_c$ may be changed by altering the hydrogen bonds, in hydrogen compounds, using a simple model of $H_2 S$. We have shown that changes in $x_0$ (half $S$ atoms separation in the model), due to  changes in pressure affect the value of the superconducting gap. This gap is optimized at a specific value of $x_0$, which is determined by $P$ (pressure of the system), since $\Delta$ decreases for too high values of $x_0$. This is both theoretically and experimentally confirmed \cite{Goncharov2017,Eremets2016}. Our model gives a qualitative explanation of the isotope effect (both normal, and inverse), depending on the value of $x_0$, or $\omega_0$ (in the H.O.). More accurate determination of the parameters of the $H-S$ potential, can, we believe, allow to fully account for $\Delta$ and its dependence on physical parameters. 
At present, very high pressures (of the order of 100 GPa) are needed to bring about the superconductivity in these compounds. However, judicious Chemistry with larger and smaller ions might create internal pressures in the material to simulate such conditions.
\section*{Acknowledgements}
Y.I. acknowledges the late Prof. I Pelah for conversations on hydrogen bonds, and the late Prof. W Kohn on superconductivity and DFT. D.K. thanks Or Ben Zvi for his immeasurable support and attention. Financial support from the Weizmann Institute of Science is gratefully acknowledged.

\end{document}